**Title**

Deposition rate controls nucleation and growth during amorphous/nanocrystalline competition in sputtered Zr-Cr thin films.

**Authors**

Q. Liebgott[a,b], A. Borroto[a], Z. Fernández-Gutiérrez[a], S. Bruyère[a], F. Mücklich[b], D. Horwat[a,]*

[a] Université de Lorraine, CNRS, IJL, F-54000 Nancy, France

[b] Department of Materials Science and Engineering, Chair of Functional Materials, Saarland University, Campus D3.3, D-66123 Saarbrücken, Germany

* Corresponding author. E-mail address: david.horwat@univ-lorraine.fr (D. Horwat)

**Abstract**

Dual-phase Zr-based thin films synthesized by magnetron co-sputtering and showing competitive growth between amorphous and crystalline phases have been reported recently. In such films, the amorphous phase grows as columns, while the crystalline phase grows as separated cone-shaped crystalline regions made of smaller crystallites. In this paper, we investigate this phenomenon and propose a model for the development of the crystalline regions during thin film growth. We evidence using X-ray diffraction (XRD), scanning electron microscopy (SEM) and transmission electron microscopy (TEM), that this competitive self-separation also exists in co-sputtered Zr-Cr thin films with Cr contents of ~84-86 at.%, corresponding to the transition between the amorphous and crystalline compositions, and in the Zr-V system. Then, to assess the sturdiness of this phenomenon, its existence and geometrical characteristics are evaluated when varying the film composition and the deposition rate. The variation of geometrical features, such as the crystalline cone angle, the size and density of crystallites, is discussed. Is it shown that a variation in the deposition rate changes the nucleation and growth kinetics of the crystallites. The surface coverage by the crystalline phase at a given thickness is also calculated for each deposition rate. Moreover, comparison is made between Zr-Cr, Zr-V, Zr-Mo and Zr-W dual-phase thin films to compare their nucleation and growth kinetics.

1. Introduction

During the last decades, interest for new functional surfaces has continuously been growing [1–3]. This is because the functionalities of surface-modified materials are manifold: antibacterial surfaces [4–6], solar cells [7], wear or corrosion protection [8–10], to name a few. Many advances in the field of thin films have been made thanks to elaboration techniques such as magnetron sputtering, allowing to inexpensively synthesize thin films at low temperatures. Among these thin films, Zr-based thin film metallic glasses (TFMGs) have been intensively investigated due to their good mechanical properties such as hardness beyond 10 GPa, improvement of fatigue resistance of 316L stainless steel, and their corrosion resistance, for instance [11–13].

The ease to synthesize such TFMGs is characterized by the glass-forming ability (GFA). The higher the GFA, the easier to sputter-deposit a TFMG. For the GFA to be the highest, the sputtered film must contain different elements, with a difference greater than 10 % in atomic radii, have a negative mixing enthalpy, and a high quenching rate [14]. Moreover, in the case of Zr-X systems (X = Cr, V, Mo, W), intermetallic phases can be found in the equilibrium phase diagram, which increase the GFA, despite the atomic radii of the different elements is rather close and the mixing enthalpy is only slightly negative (-4 to -12 kJ/mol)

[15]. This means that although it is possible to find amorphous films in a wide range of compositions, it is still possible to sputter-deposit crystalline films outside of this range. Thus, sputtering films at the transition between the amorphous and crystalline compositions could lead to new microstructures, such as the dual-phase crystalline and amorphous thin films that have been reported in the literature. This dual-phase morphology has already been observed in Zr-W [16–18] and Zr-Mo [19]. Moreover, similar morphologies have been reported in the literature for different systems (Ti-Al [20], Ti-O [21], Al-N [22]), where it has only been mentioned without much attention given to its development. These dual-phase thin films are the result of a competitive growth between the amorphous and crystalline phases during film growth, leading to unique surface morphologies [16, 19]. A possible origin for crystallization in a given range of composition is that GFA is favored, meaning that the liquid is relatively stable compared to the crystalline phase. In other words, when the composition gets closer to the crystalline composition nucleation becomes easier. Moreover, Borroto et al. have evidenced that, in the Zr-W system, films undergo evolution from amorphous, to nucleation at the column boundaries, to random nucleation as the tungsten content increases and nucleation rate increases [16].

Even if this dual-phase morphology has been observed in different binary systems, the underlying mechanisms governing the development of competitive growth of the two phases are currently not well understood. Also, the sturdiness of the phenomenon regarding the deposition conditions is still unknown, and it is unknown if this phenomenon is present regardless of the deposition conditions, such as deposition rate, or not. This is mainly due to the facts that this phenomenon has only been first observed recently, and the composition range in which it occurs is narrow, making it difficult to observe experimentally.

In this study, we show that this phenomenon can be extended to Zr-Cr co-sputtered thin films. The competitive growth is tested by changing the deposition rate between 3.8 and 74 nm/min to test if it can exist over a wide range of deposition rates, if the composition range in which it occurs changes, and to compare the geometrical features of the obtained crystalline regions. Finally, nucleation and growth evolutions are explored in order to explain our results and generalize our understanding of the competitive growth process in Zr-based sputtered thin films. It is shown that the angle of crystalline cones decreases when increasing the deposition rate, and that it is due to a difference in nucleation and growth kinetics of the crystallites inside the cones. Then, these results are compared with dual-phase Zr-V, Zr-Mo and Zr-W thin films in regard to the geometrical features of the crystalline phase.

## 2. Materials and methods

### 2.1 Thin films synthesis

Nanostructured Zr-Cr thin films were deposited onto (100) monocrystalline Si substrates (15 x 15 mm²) using DC magnetron co-sputtering of Zr and Cr metallic targets (targets dimensions: 2 inches diameter, 3 mm thickness and purity higher than 99.9 %) in an argon atmosphere. The two targets were in confocal configuration and the substrate-holder rotation was set at 15 rpm to ensure homogeneity of the deposited films. The sputtering chamber was pumped down via a mechanical and a turbo-molecular pumps allowing a base pressure of $10^{-8}$ mbar. The working argon pressure was 2 Pa, and the targets to substrate distance was set at 9 cm for the two targets. The deposition rate was controlled by the discharge current applied to the Zr target (0.04 A, 0.075 A, 0.15 A or 0.3 A). To each discharge current value corresponds one deposition rate reference (respectively 5, 11, 30 and 63 nm/min). For each deposition rate reference, the discharge current applied to the Cr target varied from 0.09 A to 0.72 A. This allowed controlling the chemical composition of the sputtered thin film, as the ratio of atoms sputtered from each target was changed with minor modification of the deposition rate around the reference rate (for more information, see Fig. S1 in Supplementary Information).

This allowed changing the composition between 82 and 88 at.%Cr for deposition rates between 3.8 and 74 nm/min. The deposition time for these films varied from 20 to 240 min, the first being for the highest deposition rate and the latter for the lowest deposition rate. This allowed to synthesize thin films with a thickness of 1-1.2 µm. For Zr-V thin films, the chamber parameters were the same as for Zr-Cr thin films. The discharge currents applied to Zr and V targets (the V target has the same dimensions as Zr and Cr targets, with a purity higher than 99.9%) were 0.1 A and 0.42 A, respectively. The deposition time was 105 min. This resulted in 2 µm thick thin films containing 86 at.%V.

For comparison purposes, a Zr-Mo and a Zr-W thin film have also been deposited. For Zr-Mo, a 3.7 µm thick thin film has been deposited in 150 min (24 nm/min deposition rate) with 0.3 A and 0.28 A applied to the Zr and Mo targets, respectively. The chamber parameters and target geometry are the same. For Zr-W, a 3.1 µm thick thin film has been deposited in 60 min (52 nm/min deposition rate) at a 3 Pa working pressure with 0.3 A and 0.5 A applied to the Zr and W targets, respectively. The W target geometry was the same as for the other targets used in this study.

During the experiments, the substrate temperature has been monitored using a thermocouple and did not exceed 45 °C for all depositions.

2.2 Thin films characterization

Film thickness was measured with a Bruker Dektak XT contact profilometer. X-ray diffraction (XRD) measurements were conducted in the Bragg-Brentano configuration with KαCu radiation (λ = 1.5406 Å) using an AXS Bruker D8 Advance diffractometer. Scanning electron micrographs were taken with a Zeiss GeminiSEM 500 scanning electron microscope (SEM) to analyze the surface morphology and the cross-section of the films. Film composition was obtained using X-ray energy dispersive spectroscopy (EDS) integrated in the SEM. Transmission electron microscopy (TEM) analysis was conducted using a cold FEG JEOL ARM200 microscope. Cross-sectional TEM samples of films were prepared using a focused ion beam (FIB) scanning electron microscope dual beam system (FEI Helios 600). Throughout the FIB process, the time during the ionic cuts was the shortest possible to avoid any heating effect.

### 2.3 Image processing

Surface coverage by the crystalline phase was obtained using a homemade image treatment MatLab program on at least 3 different cross-section SEM micrographs and analyzing it at a film thickness of 750 nm (for more information, see Fig. S3 and Fig. S4 in Supplementary material). We defined the linear coverage by the crystalline phase as the ratio between the length corresponding to the crystalline phase and the total length of a line drawn at a given thickness of the film, and we extrapolated it to be equal to the surface coverage by the crystalline phase. Cone angles and size of crystallites were determined from cross-section SEM micrographs using ImageJ software using at least 10 cones for each calculation. For the crystalline geometry results, the deposition rates will be presented as reference deposition rates, with each reference corresponding to a given applied current on the Zr target, which means that 4 points representing each applied discharge current on the Zr target will be studied. The number of crystallites per cone was obtained by calculating the ratio of the average crystalline cone volume divided by the average crystallite volume. The average cone volume was calculated using the formula $V_{cone} = \frac{\pi * h^3 * tan^2(\frac{\alpha}{2})}{3}$, where h is the height of the cone (set at 500 nm, to avoid miscalculation due to coalescence of the crystalline cones), and α the crystalline cone angle. The average crystallite volume was obtained by considering the crystallite as a cylinder; its volume was calculated using the formula $V_{crystallite} = \pi * R^2 * h$, where R is the radius of the crystallite, and h its length. The obtained ratio is the number of crystallites per crystalline cone, and was used to extrapolate the nucleation rate of crystallites inside the cones.

## 3. Results and discussion

When synthesizing Zr-Cr thin films by magnetron sputtering, a composition-driven transition from amorphous to crystalline Cr(Zr) solid solutions has been reported at ~89 at.%Cr [23]. This work aims to study what happens at the transition range, and aims to search for a competitive growth between an amorphous and crystalline phases, as has been seen for other Zr-based binary systems in magnetron sputtering [16-19]. For this purpose, different thin films at compositions around the transition between these two phases were deposited at a 11 nm/min deposition rate. X-ray diffractograms of three of these thin films are shown in Fig. 1 (a), with compositions ranging from 83 to 88 at.%Cr, close to the 89 at.%Cr reported in [23]. The film with 83 at.%Cr is totally X-Ray amorphous, and the film with 88 at.%Cr is totally crystalline. However, for a composition between these two (at 85 at.%Cr), we can see a mixture of the two XRD signals, showing that the two phases coexist in the same film. The shift in diffraction angle for the (110) bcc Cr plane is due to a change in the lattice parameter. Indeed, increasing the Zr content in the Cr(Zr) solid solution increases the lattice parameter as Zr has an atomic radius larger than that of Cr, thus making the peak to shift towards lower diffraction angles.

The same thin films have been characterized with SEM, and their surface micrographs are shown in Fig. 1 (b-d). As can be seen, the amorphous and crystalline films present different surface morphologies. At the composition of the transition, however, two different regions can be observed, an amorphous and a crystalline, separated by an interface. According to previous works [16–19], the crystalline region is in the form of disks on the surface, surrounded by an amorphous matrix. From this micrograph, it is also noticeable that, at some point, the crystalline regions coalesce. A cross-section SEM micrograph, Fig. 1 (e), shows that the amorphous phase grows in a columnar microstructure, while the crystalline phase grows as cones, with a dome on the surface. These cones do not nucleate from the beginning of deposition, but after a certain thickness, and the origin of each cone is not at the same thickness. This explains why the area of each cone emerging at the surface is different. Indeed, the surface of the crystalline region on the surface depends on the thickness where the cone nucleates, because the larger the cone, the larger its surface area. The obtained microstructure for Zr-Cr thin films is the same as what has been observed in the other Zr-based systems. Moreover, the cone angle is nearly the same whatever its nucleation thickness.

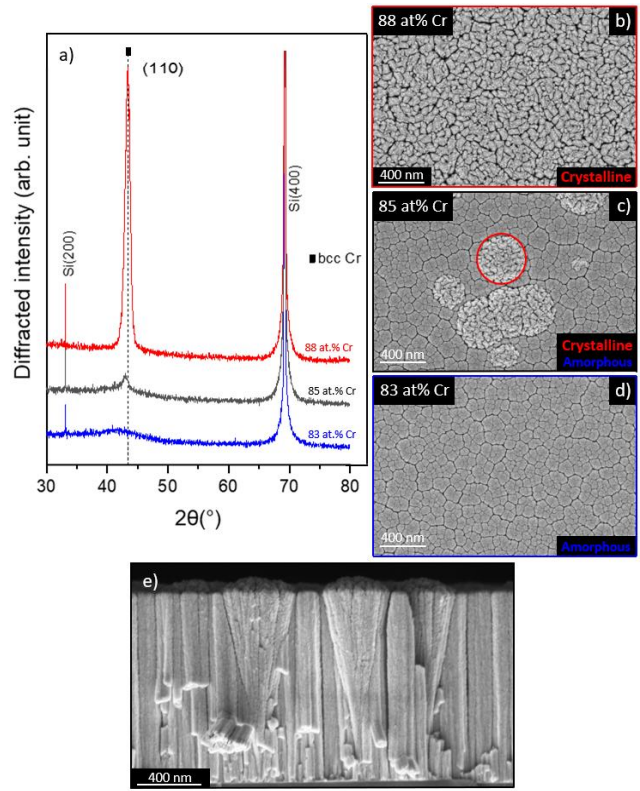

Fig. 1: (a) X-ray diffractograms of Zr-Cr thin films with 83 at.%Cr (blue), 85 at.%Cr (black), and 88 at.%Cr (red). (b-d) Surface SEM micrographs showing crystalline (b), dual-phase (c) and amorphous (d) surface morphologies, corresponding to the X-ray diffractograms on the left. (e) Cross-section SEM micrographs showing the dual-phased sample, with amorphous columns and crystalline cones with a dome shape on the surface.

Interestingly, the surface morphology can be tailored by changing the film thickness. Indeed, as thickness increases, the surface coverage by the crystalline phase increases, until it eventually covers the whole surface. Moreover, changing the composition inside the transition zone can also be used to control the surface morphology [15], as the nucleation density of the crystalline cones increases with an increasing Cr content. Therefore, combining thickness and composition can help tailoring the surface of the deposited Zr-Cr thin films. Nonetheless, the range of this competitive growth is less than 5 at.%Cr, thus very careful investigations are required to observe what happens during growth and create models.

To further understand how the competitive growth develops and can be manipulated, the deposition rate of the Zr-Cr thin films has been varied in the range 5-63 nm/min reference deposition rates. Fig. 2 shows the surface coverage by the crystalline phase as a function of the film composition and for various deposition rates. The value is measured at a 750 nm thickness for all the films. Films with a 0% surface coverage are totally amorphous, while films with a 100% surface coverage are totally crystalline at the surface, meaning they either are crystalline from the start of deposition or that crystalline cones overgrew the amorphous phase and covered the whole surface. The existence of the competitive growth is approximately in the range of 84 to 86 at.%Cr for all the deposited films. Yet, due to the narrow composition range to observe this competitive growth, it is difficult to conclude whether or not the deposition rate affects the transition composition range.

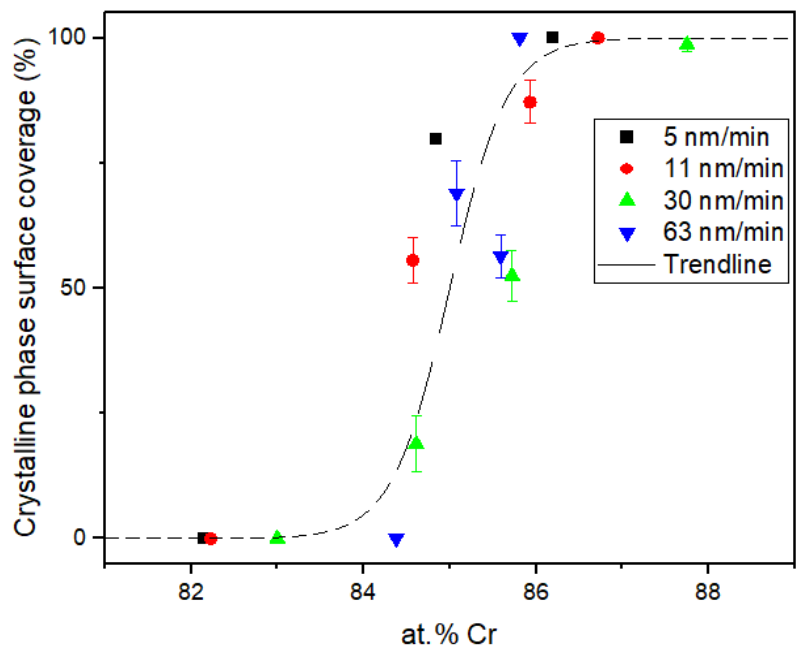

Fig. 2: Values of the surface coverage by the crystalline phase at 750 nm thickness, extrapolated from cross-section SEM micrographs, for each deposition rate range. Some error bars are too small to be seen.

However, it is important to note that in this wide range of reference deposition rates, from 5 to 63 nm/min, the competitive growth has always been observed, which indicates that it is a process resilient to changes in deposition conditions and is encouraging to find it in other systems. Yet, it is important to notice that the overall geometry of the structures is affected by the deposition rate, as shown in Fig. 3 (a-b). As can be seen, the crystalline cone angle for low deposition rates, in Fig. 3 (a), is larger than for high deposition rates, in Fig. 3 (b). Hence, a reduction of the crystalline cone angle is observed when increasing the deposition rate. Furthermore, a difference is observed depending on the alloy deposited, as can be seen when comparing Zr-Cr thin films with Zr-V thin films in Fig. 3 (c). Note that Fig. 3(c) proves that the competitive growth also exists in the Zr-V system.

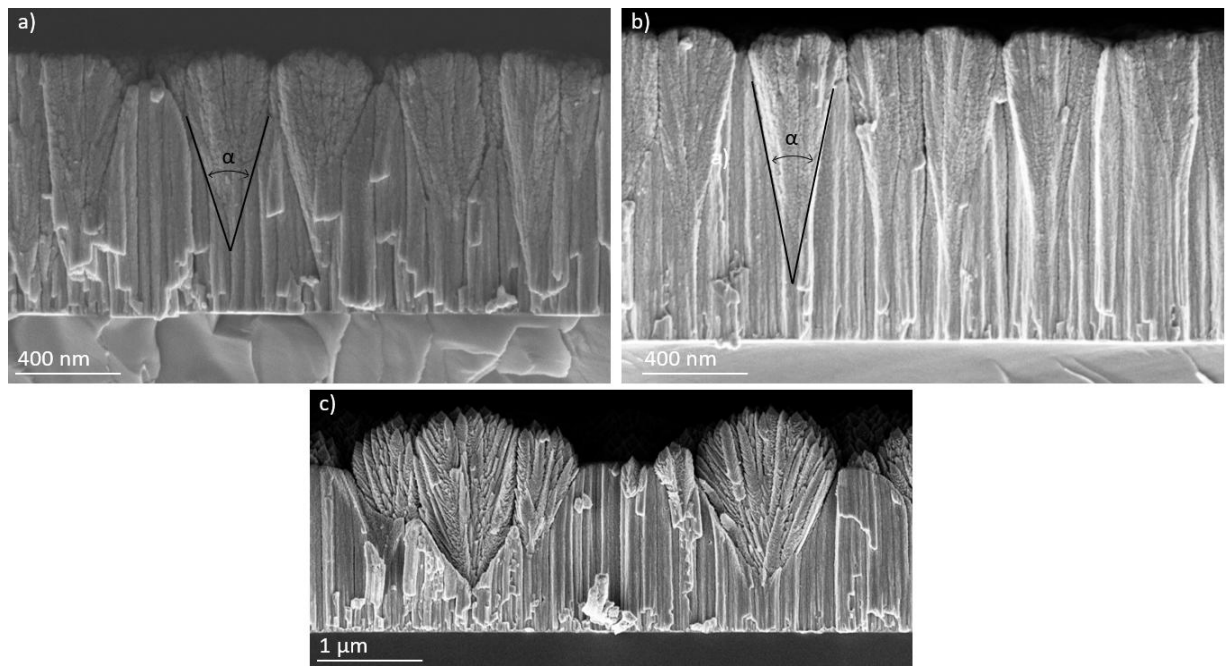

Fig. 3: SEM micrographs showing Zr-Cr thin films (~85 at.%Cr) deposited at 5nm/min (a), 63 nm/min (b), and a Zr-V thin film (86 at.%V) deposited at 18nm/min (c) for comparison.

To ensure that the columns and cones are two different phases, HRTEM and electron diffraction have been conducted on the sample presented in Fig. 1 (c). A HRTEM micrograph of a cone and its interface with the amorphous matrix is shown in Fig. 4 (a), and the selected area diffraction (SAED) patterns of the two phases are shown in Fig. 4 (b, c). These diffraction patterns confirm the crystallinity of the cones in the thin film, while the columns only present short range ordering, associated with an amorphous phase, see Fig. S2 (c) of the supplementary material. The amorphous halo of Fig. 4 (c) presents a radius that is in agreement with that of (110) bcc Cr shown in Fig. 4 (b). Therefore, we can argue that the local bcc Cr order present in the amorphous columns is the precursor of the nucleation of crystalline regions.

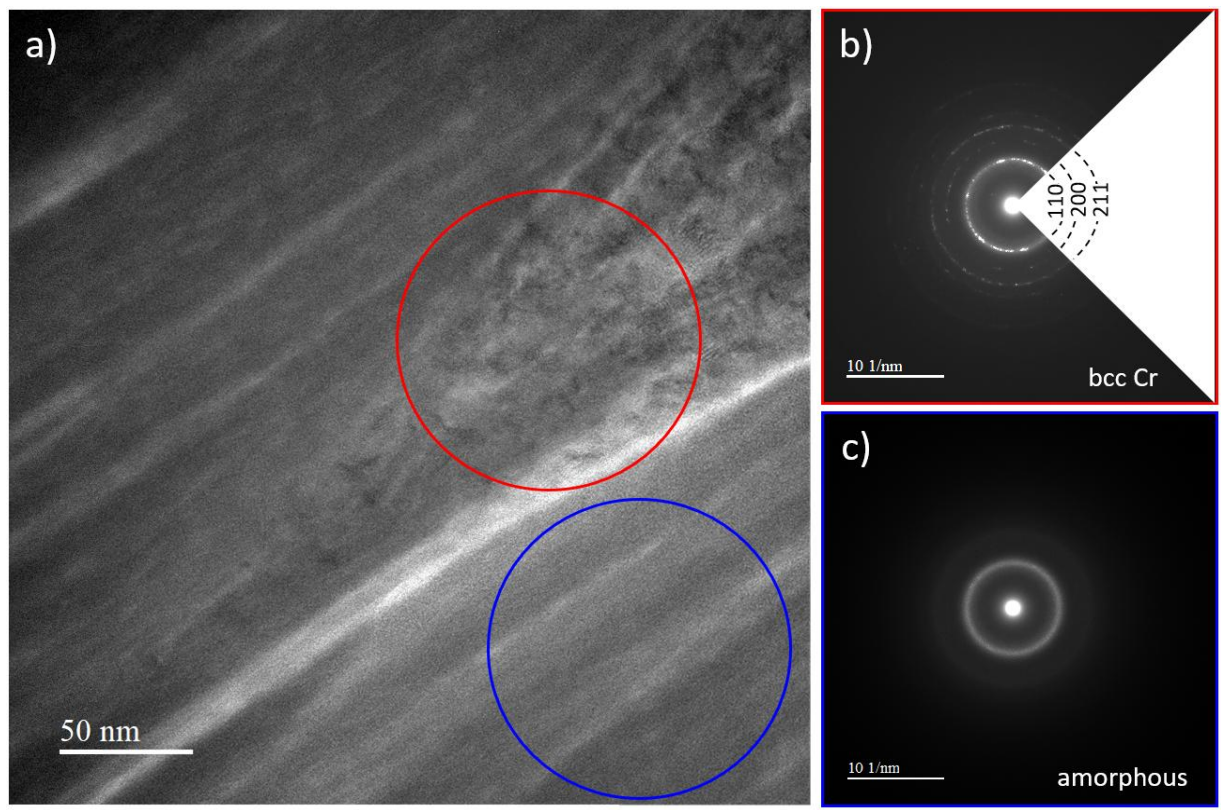

Fig. 4: HRTEM micrograph (a) showing the interface between a cone (in the middle) and a column (surrounding the cone), and (b) and (c) the selected area diffraction (SAED) patterns of the zones highlighted in (a).

As these films showing competitive growth contain two distinct phases, their respective chemical composition has to be assessed in order to evaluate if the nucleation and growth of the crystalline phase could be triggered by a composition gradient. For this, EDS has been conducted on this sample inside the transmission electron microscope. The bright field TEM image in Fig. 5 (a) shows the zone where the EDS measurements have been conducted. The measurement points are shown on the blue horizontal line, and correspond to the points where the chemical composition is reported on the graph in Fig. 5 (b). As can be seen, the chemical composition change between the columns and the cone is not significant, taking into account the uncertainty associated with the EDS technique.

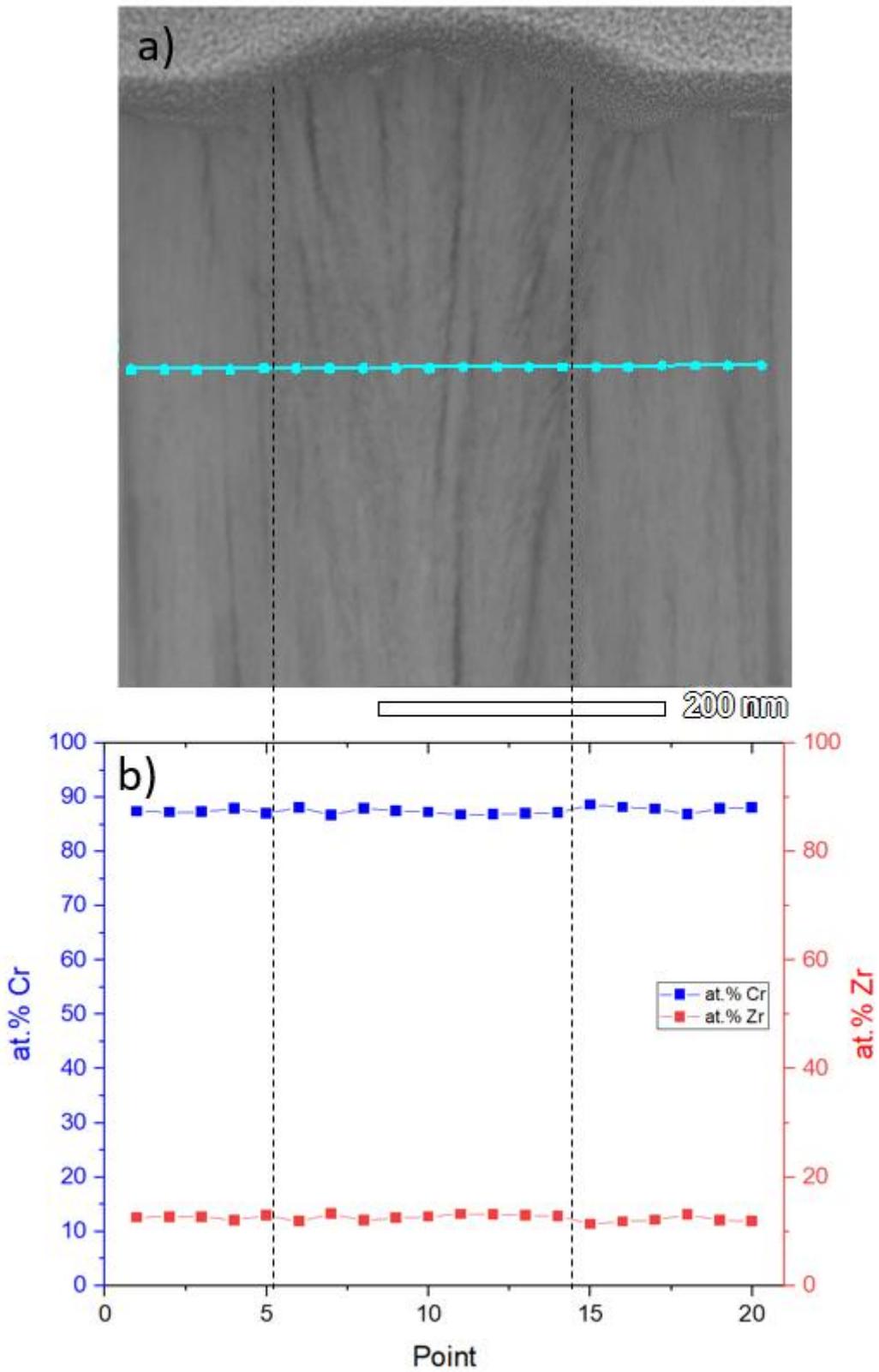

Fig. 5: Bright field TEM image showing the thin film with a crystalline cone surrounded by amorphous columns (a) and the Pt film on top of it. The points on the cyan line where the

EDS measurements have been conducted, and the composition of the film in Cr and Zr is given for each measurement point in (b). The vertical dotted lines are the intersection points between the amorphous/crystalline interface and the EDS measurement line.

For the Zr-Cr thin films, Fig. 6 (a) gives the crystalline cone angle as a function of deposition rate. When the deposition rate increases, the crystalline cone angle decreases. However, this decrease is quite small, given the fact that the deposition rate has been increased approximately 12-fold. The Zr-V thin films were deposited at a single deposition rate of 18 nm/min, resulting in the cones having an average angle of 48.4°, which is much larger than for Zr-Cr thin films. Zr-Mo thin film deposited at a 24 nm/min deposition rate resulted in a 39.9° cone angle, while the Zr-W thin film deposited at a 52 nm/min resulted in a 45.4° cone angle.

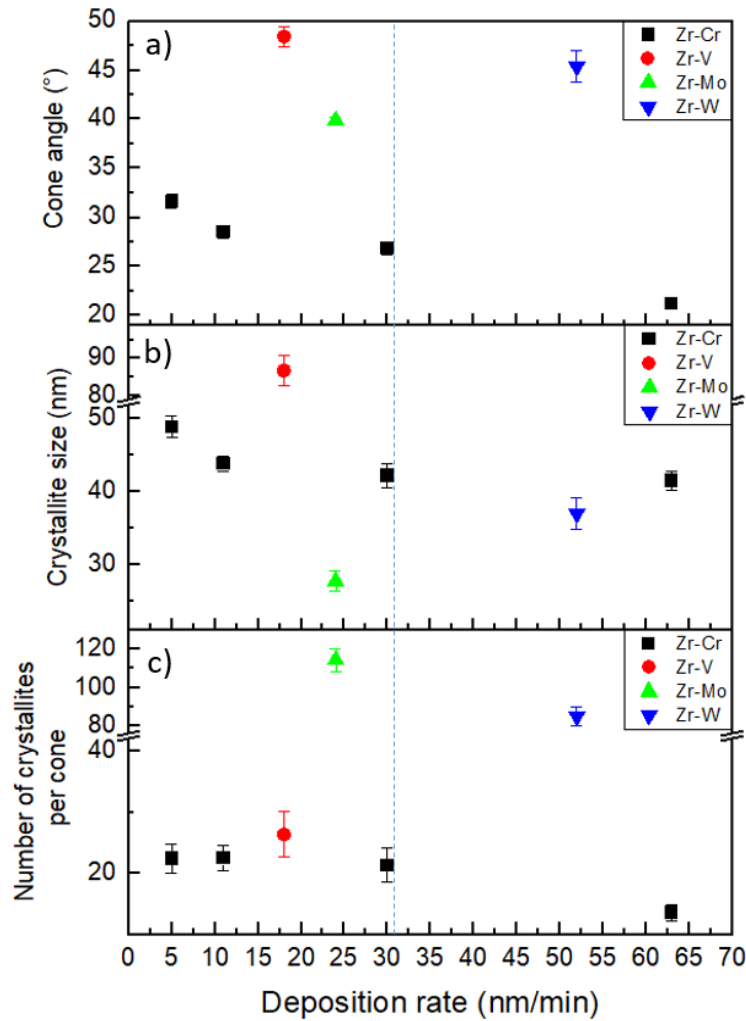

Fig. 6: Graphs showing the change in crystalline cone angle (a), in crystallite size (b) and in number of crystallites per cone (c) for Zr-Cr, Zr-V, Zr-Mo and Zr-W thin films depending on the deposition rate of the thin films. Some error bars are too small to be seen.

To try to explain the difference found depending on the deposition rate, we propose a model of nucleation and growth of crystallites inside the crystalline cones, where there is only a small number of crystallites that nucleate at the apex of the cone, and the other crystallites nucleate at the interface between the growing crystallites and the amorphous columns upon the development of the crystalline cones. The proposed model is shown in Fig. 7.

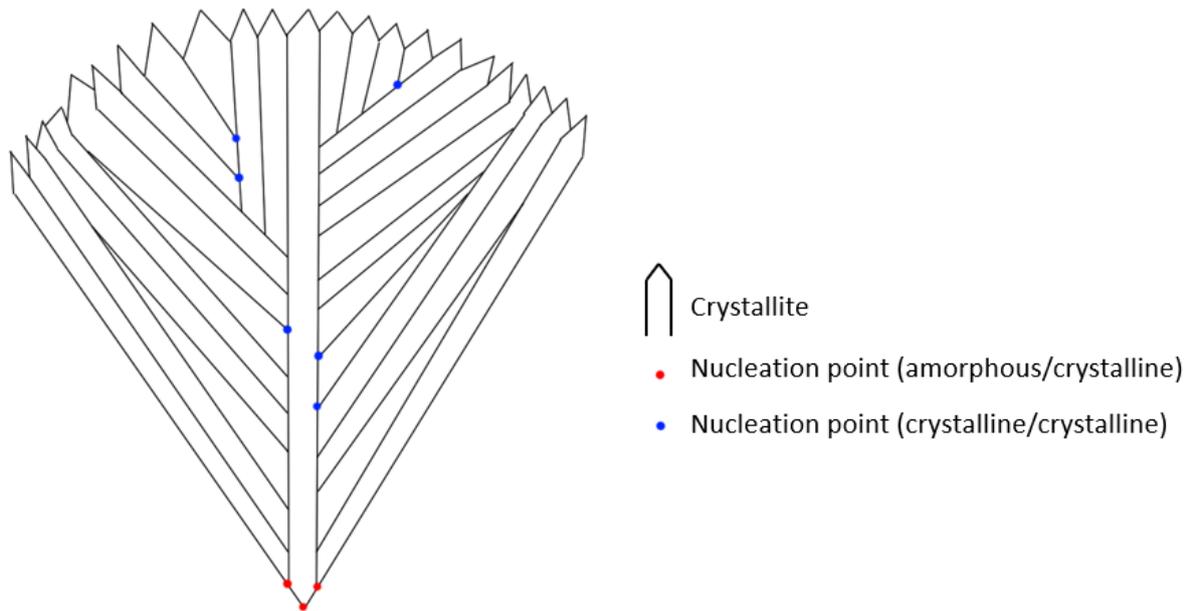

Fig. 7: Proposed nucleation and growth model for the crystalline cones in Zr-based thin films.

For the nucleation of a crystalline cone to occur, an energy barrier has to be overcome. The energy needed can mainly be obtained chemically, thermally or mechanically. As suggested by the EDS results on the TEM sample, the composition regarding Zr and Cr elements is the same throughout the whole sample, so it is unlikely that a local chemical gradient is at the origin of the nucleation. Moreover, during the whole deposition, the substrate temperature did not exceed 45 °C, which makes the hypothesis of thermal energy being the precursor of the nucleation less plausible. The mechanical energy originates from local changes of the strain energy due to an uneven distribution of the stress through the film. It has been shown that these dual-phase thin films exhibit a change of internal mechanical stress at the first stages of the crystallite nucleation. Thus, it is likely that the origin for the nucleation of the first crystallite lies in the mechanical energy aquired thanks to mechanical stress [18].

It can be seen on the SEM micrographs (Fig. 3) that some crystallites grow from the beginning of the cone, and some other nucleate at the interface between a crystallite and the amorphous matrix, or between crystallites. It is easier to observe on the Zr-V thin film, as the crystalline phase is less dense and the size of the crystallites is larger than for Zr-Cr. On Fig. 3 (c), on the cone on the right of the micrograph, vertical crystallites can be seen at the axis of the cone. In contrast, crystallites that nucleated at the interface between the first crystallites and the amorphous matrix on the right, and on the left some crystallites show a larger angle, indicating that they nucleated higher in the cone between different crystallites.

Using this model, the change in crystalline cone angle with deposition rate might find its origin either in a difference in nucleation rate of the crystallites inside the cones, a difference in their growth kinetics, or both.

The first explanation for the decrease of the crystalline cones angle with increasing deposition rate is the surface diffusion of adatoms. Indeed, an increased deposition rate means less time for the adatoms to diffuse, and thus less in-plane growth of the crystalline cones, reducing the cone angle. Nevertheless, given the high increase in deposition rate, the in-plane growth rate would need to be 8 times higher if the diffusion was the only parameter affecting the cone angle, following the equation given by Borroto et al. [16]:

$$\frac{V_G}{V_0} = \tan\left(\frac{\alpha}{2}\right)$$

With $V_G$ the in-plane growth rate, $V_0$ the deposition rate, and $\alpha$ the crystalline cone angle, as shown on Fig. 3.

Hence, changing the deposition rate affects the in-plane growth of the crystallites. This is confirmed in Fig. 6 (b), where a ~15 % decrease in crystallite size is observed when increasing the deposition rate from 5 to 63nm/min, due to the fact that the adatoms have less time to diffuse on the surface of the film when the deposition rate is higher, and that means less in-plane growth for the crystallites.

However, this ~15 % decrease in crystallite size alone does not explain the ~30 % decrease in cone angle. According to our model, it means that there could be another phenomenon happening, such as a difference in nucleation rate of the crystallites inside the cones. Thus, if this model of nucleation and growth of the crystalline phase is correct, a decrease of the nucleation rate should be observed when increasing the deposition rate.

To quantify the difference in nucleation inside the cones, the number of crystallites in a cone must be known. Yet, two problems arise when counting the number of crystallites per cone. The first is due to the density of the crystallites in the cones, making it difficult to count them. Some interfaces can be easily seen, and have been used to calculate the average crystallite size for each deposition rate range, but it is not the case for the vast majority of the crystallites, where it becomes difficult to see the separation. The second problem is that when observing a cross-section SEM micrograph, the cones are not always cut in their center, and it is of great difficulty to know exactly where a cone has been cut, which means counting the crystallites from SEM micrographs is misleading.

To overcome these issues, the average number of crystallites per cone will be calculated by dividing the average cone volume by the average crystallite volume. To calculate the first one, the volume is calculated by using the cone angle and setting the height to 500 nm, which means for cones that have not already coalesced. For the latter, the crystallite is assumed to be a cylinder, of radius half of the crystallite size shown in Fig. 6 (b), and of length half of the 500 nm height used for the cone, as all crystallites don't nucleate at the origin of the cone. Then, the average cone volume is divided by the average crystallite volume, giving the plot in Fig. 6 (c). As can be seen, for a deposition rate between 5 and 30 nm/min, the nucleation density in a cone does not significantly change. However, when depositing films at 63nm/min, a 40 % reduction of number of crystallites is observed, which means that the nucleation rate is lower for this deposition rate, contributing to reducing the crystalline cone angle.

It should be noted that this calculation gives the same results no matter the length of the crystallites as compared to the cone height. As can be seen in Fig. 6, when increasing the deposition rate from 5 nm/min up to 30 nm/min, the decrease in crystalline cone angle is mostly due to the reduction of the crystallite size, and the nucleation density does not significantly change. Hence, at these deposition rates, the change in crystalline cone angle is mostly governed by the diffusion of the adatoms on the surface of the film during growth, whereas increasing the deposition rate from 30 to 63 nm/min did not significantly change the crystallite size. Instead, the number of crystallites per cone decreased by nearly 40 % as compared to other deposition rates, which means that the decrease in crystallite cone angle for very high deposition rates is mostly due to a lower nucleation density of the crystallites in the cones.

As a comparison, for the Zr-V thin films, as can be seen in Fig. 6 (a), the crystalline cone angle is much higher than that of Zr-Cr thin films, being 48° and from 21° to 32°, respectively. Moreover, it can be seen in Fig. 6 (b) that the crystallite size is 87 nm, approximately 2 times higher than that of Zr-Cr films, and the number of crystallites per cone is nearly the same as for Zr-Cr thin films as Fig. 6 (c) shows. Thus, the higher cone angle in Zr-V dual phased thin films as compared to Zr-Cr thin films might find its origin in the higher surface diffusion of V adatoms as compared to Cr adatoms, leading to higher in-plane growth of the crystallites in Zr-V crystalline cones and thus a higher cone angle.

An explanation for the higher surface diffusion of V adatoms as compared to Cr adatoms in the studied Zr-V and Zr-Cr alloys lies in their different melting temperatures and solidification intervals. According to the literature, the stable equilibrium peritectic temperature of our Zr-V thin films is 1573 K [24] and solidification interval is close to 400 K at the composition range of interest, while for the Zr-Cr thin films the peritectic temperature is 1832 K and the solidification interval is close to 100 K [25]. During sputtering, the surface diffusion directly depends on the ratio between the deposition temperature and the melting temperature of the deposited material [26]. At a similar deposition temperature, a lower melting temperature means a higher adatom surface diffusion. Thus, the adatom diffusion for the Zr-V system might be higher than for Zr-Cr, allowing a higher in-plane growth of the crystallites for the Zr-V crystalline cones, explaining the higher cone angle as compared to Zr-Cr for the same deposition rate. Similarly, higher solidification interval means more time for the adatoms to diffuse during the solidification process occurring at the film surface. The Zr-W peritectic temperature is 2483 K, according to its phase diagram [27], and the crystallites are smaller than in Zr-Cr and Zr-V. However, the Zr-Mo peritectic temperature is 2194 K [28], but its crystallites are even smaller than for Zr-W. Again, the potential reason for the lower diffusion length of in Zr-Mo is the smaller solidification interval (approximately 300 K compared to 700 K for Zr-W), leading to a faster solidification, which allows less time for diffusion.

4. **Conclusion**

To summarize, the competitive growth phenomenon that occurs between amorphous and crystalline phases that has been observed in the literature for sputtered Zr-Mo, Zr-W and Zr-V films has been also found in the Zr-Cr system. This study shows that it exists in a wide range of deposition rate, indicating its resilience against changes of deposition conditions and giving promises for its experimental observation in other systems. However, changing the deposition rate also changes the geometrical features of the crystalline phase such as the cone angle, which can be used as a new parameter to control the growth of the film, together with the film composition and film thickness. The size and density of crystallites inside a cone also depend on the deposition rate and the material deposited. Furthermore, we proposed a model of nucleation and growth for the crystallites inside the crystalline cones to explain the development of this competitive growth, which is a step towards the understanding of the mechanisms underlying this process. This allows a fine tuning of the surface morphology, and hence of the functional properties. We showed that depending on the deposition rate, the change in the crystalline cone angle can be due to either the nucleation kinetics of the crystallites, or their lateral growth.

**Acknowledgements**

The "Université Franco-Allemande" (UFA) and the "Ministère de l'Enseignement Supérieur et de la Recherche" are deeply acknowledged for the PhD scholarship of Quentin Liebgott within the PhD-track in Materials Science and Engineering at UFA.